\documentclass[nofootinbib,twocolumn,aps,prl,amsmath,showpacs]{revtex4-1}
\pdfoutput=1
\usepackage{graphicx,dcolumn,bm,amssymb,amsmath}
\usepackage{xcolor}

\usepackage{float}

\usepackage{hyperref}
\def\be{\begin{equation}}
\def\ee{\end{equation}}

\usepackage{tikz}
\PassOptionsToPackage{svgnames}{xcolor}
\usepackage{pgfornament}
\newcommand{\sectionlinetwo}[2]{%
  \nointerlineskip \vspace{.5\baselineskip}\hspace{\fill}
  {\color{#1}
    \resizebox{0.5\linewidth}{2ex}
    {{%
    {\begin{tikzpicture}
    \node  (C) at (0,0) {};
    \node (D) at (9,0) {};
    \path (C) to [ornament=#2] (D);
    \end{tikzpicture}}}}}%
    \hspace{\fill}
    \par\nointerlineskip \vspace{.5\baselineskip}
  }

\begin{document}
\title{\Large Unified theory of vibrational spectra in amorphous materials}
\author{M. Baggioli$^{1}$ and   A. Zaccone$^{2,3,4}$}

\vspace{0.5cm}

\affiliation{${}^1$Instituto de Fisica Teorica UAM/CSIC, c/Nicolas Cabrera 13-15,
Universidad Autonoma de Madrid, Cantoblanco, 28049 Madrid, Spain.}
\affiliation{${}^2$Department of Chemical Engineering and Biotechnology,
University of Cambridge, Philippa Fawcett Drive, CB30AS Cambridge, U.K.}
\affiliation{${}^3$Cavendish Laboratory, University of Cambridge, JJ Thomson
Avenue, CB30HE Cambridge, U.K.}
\affiliation{${}^4$Department of Physics ''A. Pontremoli", University of Milan, via Celoria 16, 20133 Milan, Italy.}

\begin{abstract}
The phonon spectra of solids, described through the measurable vibrational density of states (VDOS), provide a wealth of information about the underlying atomic structure and bonding, and they determine fundamental macroscopic properties such as specific heat, thermal conductivity and superconductivity. In amorphous solids, structural disorder generates a number of poorly understood phenomena that are absent in crystals. Starting from an effective field theory description which accounts for disorder in terms of momentum diffusion, we provide a unified theoretical description of the VDOS of amorphous materials. The theory provides an excellent description of the simulated/experimental VDOS of two quite different materials (silica glass and amorphous Si) with very few non-trivial fitting parameters. In particular, the phonons speeds and diffusive linewidths obtained from the fits are fully consistent with the tabulated values from experiments. This semi-analytical approach solves the long-standing problem of describing and explaining the features of the VDOS and Raman spectra of amorphous solids in relation with the underlying acoustic dispersion relations and acoustic attenuation. Furthermore, it can be used in a  reverse-engineered way to estimate phonon dispersion relations, acoustic attenuation and dynamic structure factors from a simple fitting of the VDOS with the presented model.
%The analysis reveals that the boson peak and the (lowest) smoothed van Hove singularity are distinct phenomena, and that phonons are harmonically damped by the structural disorder.

\end{abstract}

\pacs{}

\maketitle

The phonon spectra of solids are described through the vibrational density of states (VDOS), which serves as  the cornerstone for the modeling of the thermal properties of solids, such as the specific heat, the thermal conductivity and also for the determination of the electron-phonon behaviour, superconducting critical temperature, etc. In crystals, the VDOS $g(\omega)$ can be calculated in a precise and straightforward manner provided that the following information is available: (i) the dispersion relations of the phonons, (ii) the pseudo-Brillouin zone (BZ) structure in reciprocal space. With this information at hand, the Van Hove singularities in the VDOS can be located, branch by branch, in a straightforward way and the whole spectrum can be reconstructed~\cite{Phillips1956}. The low-frequency part of the spectrum is dominated by the Debye law $\sim \omega^{2}$, which describes the VDOS of acoustic phonons arising from the breaking of translation symmetry \cite{Leutwyler:1996er} and following a linear dispersion relation $\omega= v \,k$. 

In amorphous materials, such as glasses, phonons behave like extended plane waves (ballistic phonons or \textit{propagons}) only on very large length scales or very low frequencies, at which their density of states is still described by Debye's law. Scattering, induced by the disorder, leads to the appearance of quasi-localized excitations (\textit{diffusons}) with a (harmonic) diffusive-like behaviour controlled by elastic scattering. The \textit{diffusons} dominate the VDOS starting from a length scale at which the excitation wavelength becomes comparable with the mean free path, the so-called Ioffe-Regel crossover. It has been shown, by Allen and Feldman~\cite{AllenFeldman,allen1999diffusons}, that \textit{diffusons} provide an effective, and very successful, way to describe disorder-induced features of heat transfer and thermal conductivity in amorphous solids, in good agreement with the famous experimental observations \cite{PhysRevB.4.2029}. 

The Ioffe-Regel crossover coincides, in glasses, with a peak in the normalized VDOS $g(\omega)/\omega^{2}$ referred to as the boson peak (BP) \cite{Shintani2008,PhysRevB.87.134203,PhysRevLett.96.045502}, whose temperature (in)dependence clearly indicates that is governed by harmonic disorder. Theories based on spatial heterogeneity of the shear elastic constant~\cite{Ruocco2007} have been able to successfully describe the boson peak and the associated crossover in sound attenuation from $\Gamma \sim \omega^{4}$ (Rayleigh) to $\Gamma \sim \omega^{2}$ at the boson peak, which has been observed experimentally~\cite{Baldi2010}. While \textit{diffusons} occupy most part of the central range of excitations in the VDOS, at the highest frequencies approaching the Debye frequency, the excitations become fully (Anderson) localized (\textit{locons}).

Furthermore, the BZ structure in glasses is less well defined due to the disorder. One should, in fact, refer to pseudo-BZs, which result in shifted and substantially smoothed van Hove peaks ~\cite{Taraskin}.
In spite of much progress in understanding these features (boson peak, smoothed VH peaks, etc), no unified theoretical framework is currently able to describe the VDOS of real amorphous materials over the entire spectrum. Here, we develop a unified theory at the level of effective field theory which for the first time is able to provide a description of the VDOS of real amorphous materials and provide a number of new insights into the physics of vibrational excitations in glasses. 

Recent numerical progress \cite{Tanaka,Parshin2} has highlighted that the linewidth $\Gamma$ (extracted from the dynamic structure factor $S(k,\omega)$) of monoatomic glasses at low $T$ goes quadratically with the wave-vector, $\Gamma \sim k^{2}$, over a broad range of $k$ across the BP. This dependence unveils the diffusive-like nature of the vibrational degrees of freedom in the \textit{diffusons} regime, with a temperature-independent diffusion coefficient which depends on the strength of harmonic disorder. 

We therefore start from an equation of motion for the elastic displacement field $\mathbf{u}$ in an isotropic solid (i.e. such that longitudinal and transverse components can be decoupled) which, besides the standard elastodynamic part coming from linear elasticity theory, contains a term describing momentum diffusion. In this way, the time evolution of the displacement field can cover both the ballistic (\textit{propagon}) and the diffusive (\textit{diffuson}) regimes, with the latter becoming comparatively more prominent upon increasing $k$, close to the Ioffe-Regel crossover. The full dynamics is then described by the simple differential equation:

\begin{equation}
\rho\,\frac{\partial^2\mathbf{u}_{\lambda}}{\partial t^2}=\rho\, v_{\lambda}^2 \bigtriangleup\mathbf{u}_{\lambda}+ D_{\lambda}\frac{\partial\bigtriangleup\mathbf{u}_{\lambda}}{\partial t},\label{unob}
\end{equation}
where $v_\lambda$ and $D_\lambda$ are respectively the speed of propagation and the diffusion constant of the $\lambda$ branch. Here the subscript $\lambda$ refers to either longitudinal $\lambda=L$ or transverse $\lambda=T$ displacement field.
Upon Fourier transforming the displacement field $\mathbf{u}$, one readily obtains the associated Green function,
\begin{equation}
    G_\lambda(\omega,k)\,=\,\frac{1}{\omega^2\,-\,\Omega_{\lambda}^{2}(k)\,+\,i\,\omega\,\Gamma_{\lambda}(k)}
\end{equation}
where the propagating term is naturally given by $\Omega_{\lambda}^{2}= v_{\lambda}^2\,k^2\,$. The diffusive damping due to harmonic disorder-induced scattering, following from Eq.\eqref{unob}, is given by $\Gamma_{\lambda}(k)=D_\lambda\,k^2$. This diffusive form of the damping is justified by several simulation studies \cite{Tanaka,Parshin1,Parshin2,Schirmacher} over a broad range of $k$. The \textit{diffusons} are also possibly related to the random-matrix structure of the eigenvalue spectrum of the dynamical (Hessian) matrix, as suggested in Refs.~\cite{BaggioliPRR,2019arXiv190305237B}. 

\begin{figure}
    \centering
    \includegraphics[width=0.9\linewidth]{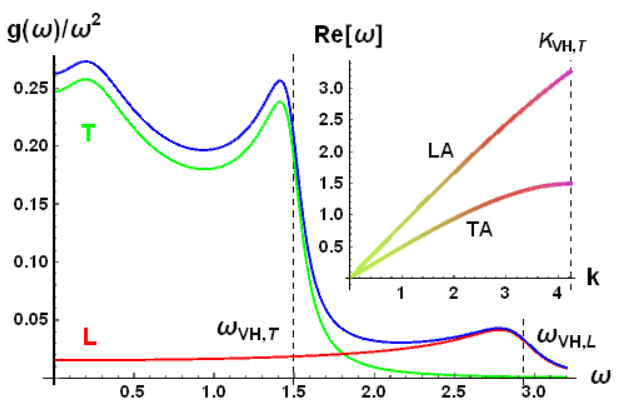}
    \caption{Calculated VDOS using the theoretical model described in the text with typical values of parameters for amorphous materials. We set $v_T=0.5,v_L=0.7,A_{T,L}=0.007,D_{T,L}=0.012,k_D=5.95$. Inset: acoustic dispersion relations used for the calculation of the VDOS. We set $v_T=0.5,v_L=0.85,A_{T,L}=0.007$.}
    \label{fig:uno}
\end{figure}

In order to be able to capture the behaviour upon approaching the pseudo-BZ boundary, we add an additional higher order term in the acoustic dispersion relations \cite{kosevich2005crystal}:
\begin{equation}
\Omega_{\lambda}^{2}=v_{\lambda}^2\,k^2\,-\,A_\lambda\,k^4\label{treEQ}
\end{equation}
such that the BZ boundary, and the corresponding Van-Hove (VH) singularity, are located at $k_{VH,\lambda}\equiv v_\lambda/\sqrt{2\,A_\lambda}$ (see inset in fig.\ref{fig:uno}). 
This expression can also be obtained from a nonlinear elastodynamic equation of the Boussinesq type~\cite{kosevich2005crystal}. In the present context, the quartic term in $k$ is needed in order to provide an accurate modelling of the flattening of the dispersion relations upon approaching the pseudo-BZ boundary.

With this important improvement, we arrive at the following final form of the Green functions:
\begin{equation}
    G_\lambda(\omega,k)\,=\,\frac{1}{\omega^2\,-\,v_{\lambda}^2\,k^2\,+\,A_\lambda\,k^4\,+\,i\,\omega\,D_\lambda\,k^2}.
\end{equation}

This Green's function takes into account the following physical aspects: (i) propagative (acoustic) phonons; (ii) diffusive-like propagation at larger $k$ due to harmonic disorder through the imaginary term; (iii) the pseudo-BZ boundaries for the various branches through the $A_{\lambda}$ coefficients. 

In a comparison with experimental/simulation data, the coefficients $D_{\lambda}$ are the only non-trivial fitting parameters, since the values of $v_{\lambda}$ and $A_{\lambda}$ need to qualitatively comply with the trend of the dispersion relations for glasses~\cite{Ruzicka}.

In Fig.\ref{fig:uno} the schematic dispersion relations computed using Eq.\eqref{treEQ} with typical parameters for amorphous materials are shown. The flattening at the pseudo-BZ boundary (where the group velocity $d\omega/dk=0$) is reproduced thanks to the quartic term. In amorphous solids, typically the transverse dispersion relation is lower and reaches the plateau much earlier than the longitudinal one~\cite{Ruzicka}, which implies a different position of the VH  peak for L and T phonons. 

The above Green's function can be used to evaluate the VDOS according to the formula
\begin{equation}
    g(\omega)\,=\,-\,\frac{2\omega}{\pi  k_D^3}\int_0^{k_D}\text{Im} \left[\,2\, G_{T}(\omega,k)+ G_{L}(\omega,k)\right]k^2 dk
\end{equation}
See the Supplementary Info of Ref.~\cite{baggioli2018universal} for a derivation of this formula.

Using typical values that qualitatively reproduce the acoustic dispersion curves in amorphous solids~\cite{Ruzicka}, the VDOS spectrum is obtained as plotted in Fig.\ref{fig:uno}, normalized by the Debye $\sim \omega^{2}$ law. It is clearly seen that a boson peak arises, similarly to what predicted in \cite{baggioli2018universal}, at the crossover from \textit{propagons} to \textit{diffusons}, in the low-frequency part of the spectrum. At higher frequencies the two VH peaks, for transverse ($\omega_{VH,T}$) and longitudinal ($\omega_{VH,L}$) phonons, are observed as a result of the flattening of the dispersion relations in \eqref{treEQ} induced by the new parameter $A_{T,L}$. In amorphous materials with strong structural disorder optical modes are not generally observed~\cite{Parshin2}. Given the hierarchy of propagation speeds $v_L \gg v_T$, the features at low frequency, and especially the BP, are totally dominated by the transverse branch of modes as confirmed in Fig.\eqref{fig:uno}. As a consequence, we consider a simplified version of the above model, where we neglect the longitudinal modes, since they provide only subleading effects at low frequencies. This allows us to better analyze the effect of the mode diffusivity $D_{T}$ on the boson peak and on the (transverse) VH peak.

\begin{figure}
    \centering
    \includegraphics[width=0.8\linewidth]{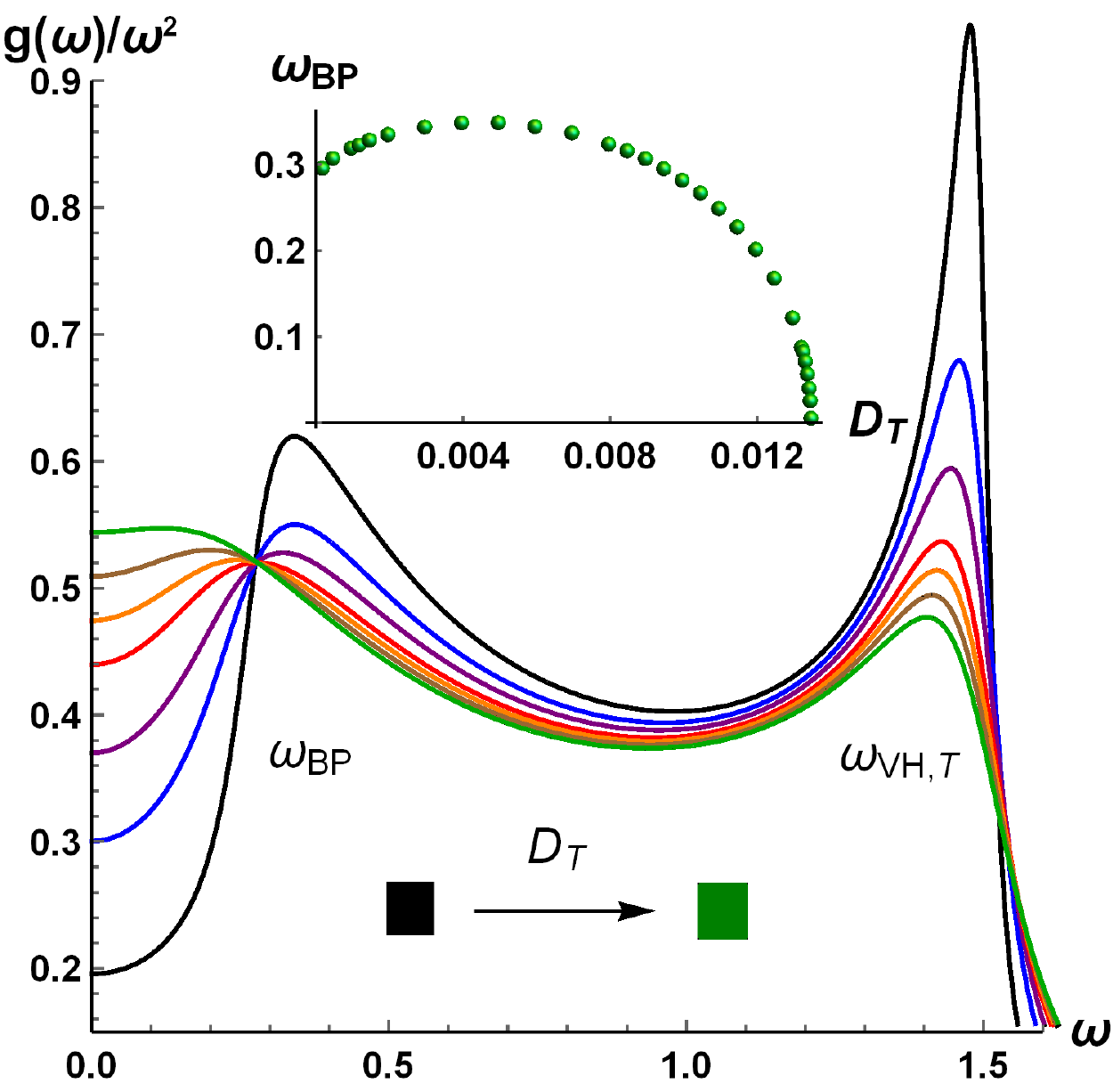}
    \caption{Effect of increasing the transverse (disorder-induced, harmonic) diffusive damping coefficient $D_T$ on the boson peak and on the lowest VH peak. Inset: dependence of the boson peak frequency on the transverse diffusive coefficient $D_T$. We set $v_T=0.5,v_L=0.7,A_{T,L}=0.007,D_L=0.012,k_D=5.95$ and we dial $D_T \in [0.003,0.013]$.}
    \label{fig:due}
\end{figure}
The calculation in Fig.\ref{fig:due} clearly shows that increasing the diffusivity $D_{T}$ -- a measure of the structural disorder in the material -- causes a broadening of both the boson peak and the VH peak. This effect is physically meaningful since disorder is known to cause a broadening of the VH peaks~\cite{Taraskin} and also to cause the predicted shift of the boson peak towards lower frequency (ultimately the boson peak approaches $\omega=0$ at ''epitome of disorder", i.e. in marginally stable jammed packings at the verge of rigidity~\cite{Silbert}). In the dependence of the boson peak frequency upon $D$ we note also an interesting non-monotonicity as shown in the inset of Fig.\ref{fig:due}. It is important to note that without the diffusive damping $D_{T}$ the boson peak does not arise, and instead a Debye $\sim \omega^{2}$ trend would be observed up to the first VH singularity.

\begin{figure}[b]
    \centering
    \includegraphics[width=0.8 \linewidth]{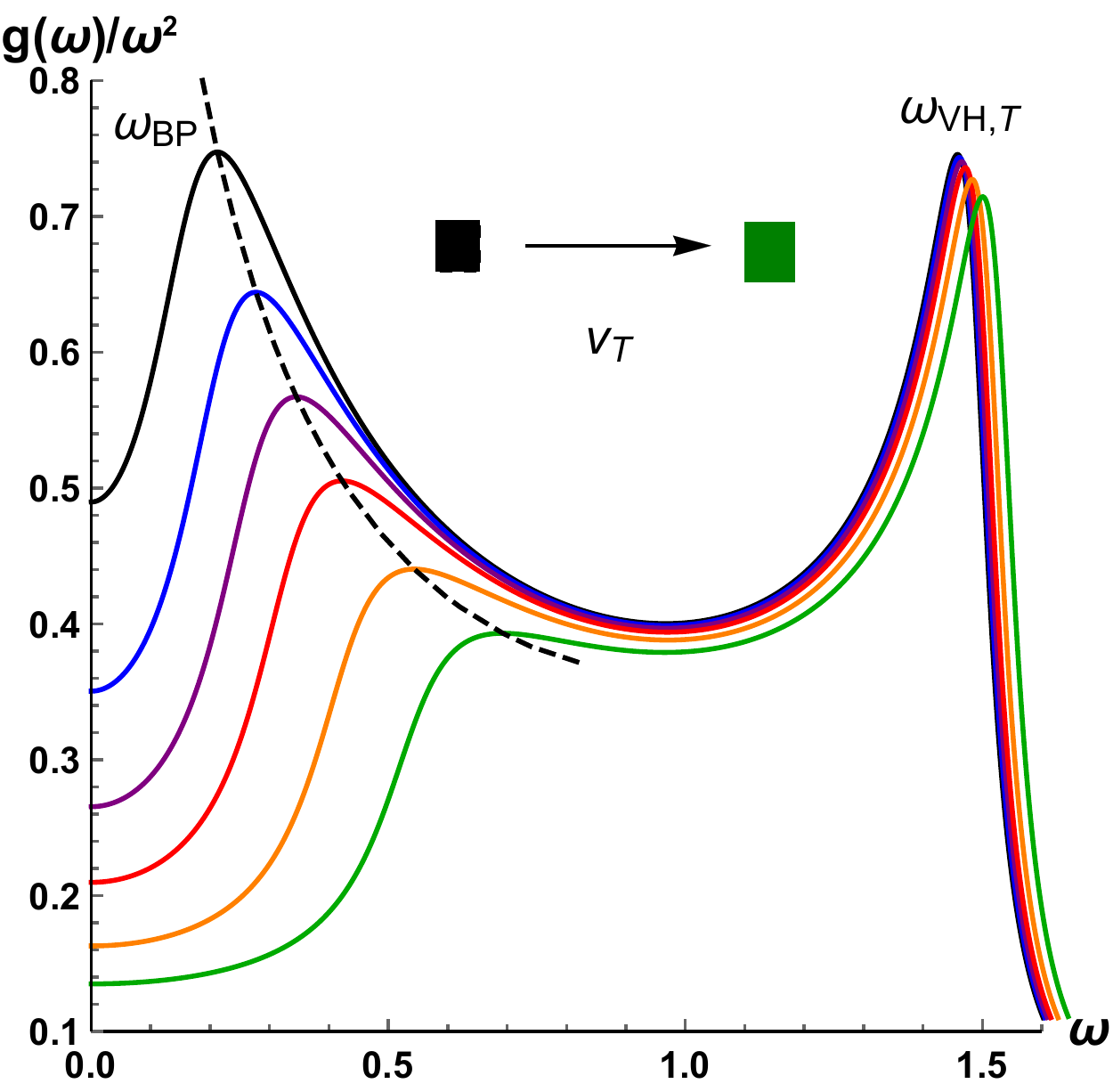}
    
    \vspace{0.5cm}
    
    \includegraphics[width=0.8 \linewidth]{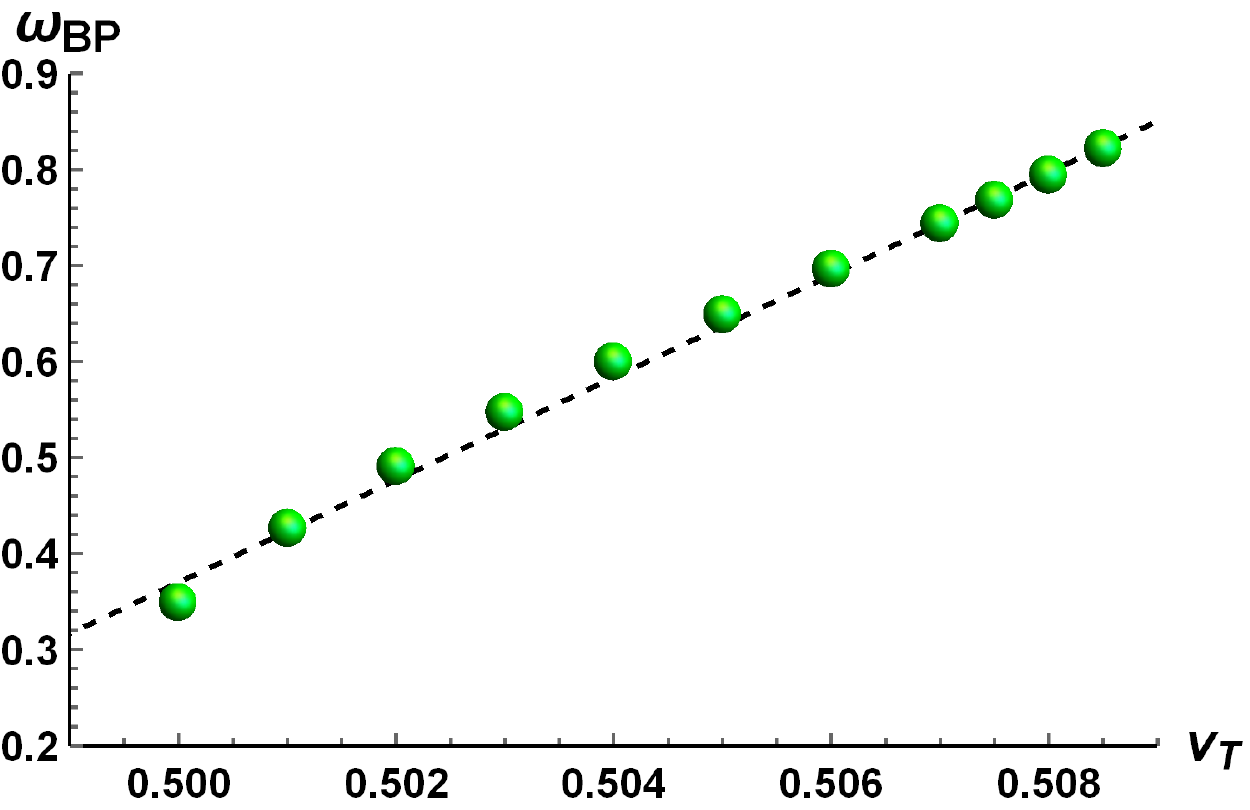}
    \caption{\textbf{Top}: Effect of increasing the (transverse) speed of sound $v_{T}=\sqrt{G/\rho}$, where $G$ is the shear modulus and $\rho$ is the density, on the VDOS (using the simplified model with only transverse modes). We set $v_L=0.7,A_{T,L}=0.007,D_{T,L}=0.012,k_D=5.95$ and we dial $v_T \in [0.4988,0.506]$. \textbf{Bottom}: the linear dependence of the boson peak frequency $\omega_{BP}$ on the speed of sound predicted by our theory. A linear dependence implies a square-root dependence of $\omega_{BP}$ on the shear modulus $G$.}
    \label{fig:tre}
\end{figure}

In Fig.\ref{fig:tre} (top panel) the effect of increasing the speed of sound, hence the elastic modulus, is shown. The theory recovers the well known fact that an increase in the shear modulus shifts the boson peak to higher frequency, as seen e.g. in previous simulations work ~\cite{Silbert,Parshin1,Rico}.
In the bottom panel of Fig.\ref{fig:tre} the dependence of the boson peak frequency $\omega_{BP}$ upon the transverse speed of sound is shown. A linear dependence clearly emerges from the data, which, since $v_{T} \sim \sqrt{G}$, implies that $\omega_{BP} \sim \sqrt{G}$. This result has been recently observed in MD simulations of polymer glasses~\cite{Mizuno2019}. 

\begin{figure}
    \centering
    
    \includegraphics[width=0.9 \linewidth]{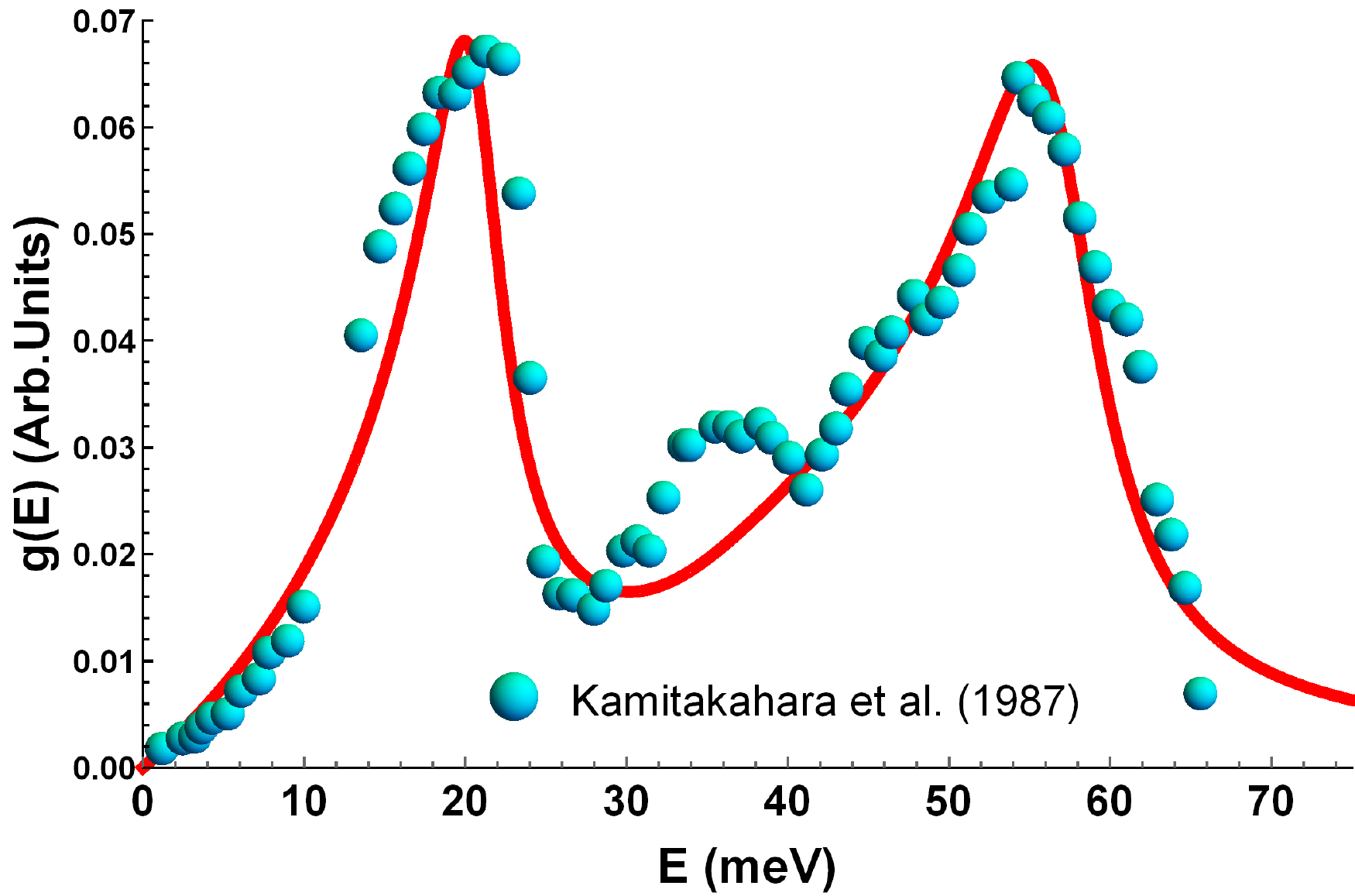}
    
    \vspace{0.5cm}
    
    \includegraphics[width=0.9 \linewidth]{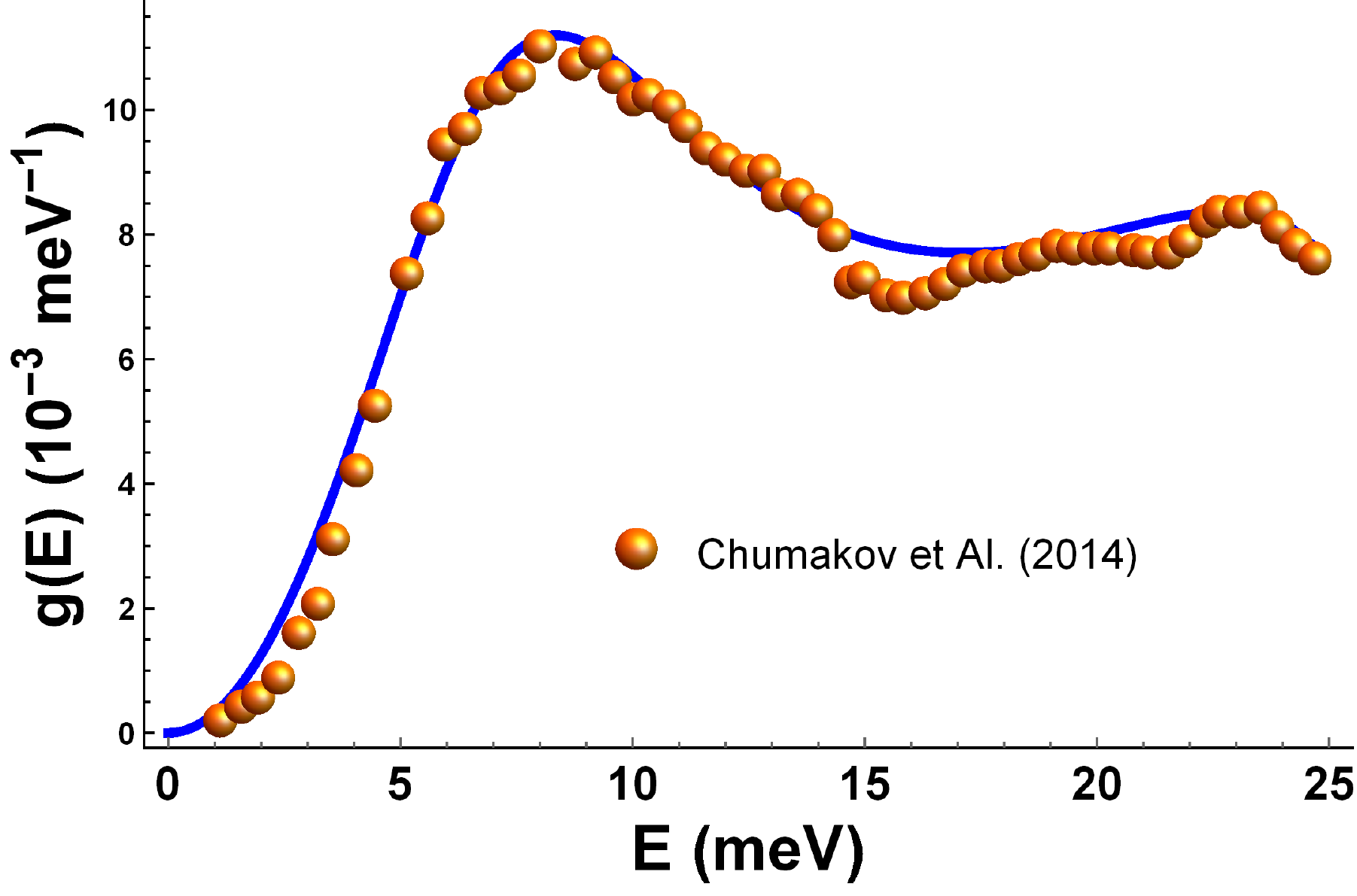}
    
    \caption{\textbf{Top: } Comparison between the VDOS spectrum calculated using our theory and the one obtained from inelastic neutron scattering ~\cite{Kamitakahara} (see legend). The ratio between the longitudinal and transverse speeds is taken as $v_L/v_T \approx 1.7$, compatible with the experimental values for the bulk and shear moduli in amorphous silicon. The concrete values of the parameters are $v_T=0.31,v_L=0.527,k_D=5.95,D_T=0.017,D_L=0.01,A_T=0.006,A_L=0.007$. \textbf{Bottom: }Comparison between the VDOS spectrum calculated using our theory, with $v_L/v_T \approx 1.8$, and experimental data for ambient silica glass from Ref.~\cite{Monaco2014} (see legend). In this plot only the lower $20\%$ of the spectrum is shown. The concrete values of the parameters are $v_T=0.4,v_L=0.72,k_D=3.3,D_T=0.133,D_L=0.03,A_T=0.009,A_L=0.007$.}   
    \label{fig:quattro}
\end{figure}

Finally, in Fig.\ref{fig:quattro} we present the comparison between the theory and the simulations and experimental data of real amorphous materials\footnote{Notice that the spectra in fig.\ref{fig:quattro} are not normalized by the Debye contribution $\sim \omega^2$. As a consequence, the BP is not apparent; the visible peaks are the VH peaks.}. In the top panel we show the comparison between our predicted spectrum for amorphous Si and the experimental ~\cite{Kamitakahara} and simulation~\cite{allen1999diffusons} results from the literature. The only non-trivial fitting parameter in the model is the diffusion coefficient of the modes ($D_{L,T}$), whereas the other parameters are constrained to reproduce plausible shapes of the acoustic dispersion relations~\cite{Parshin2,Ruzicka}. The theoretical description provides an excellent fit to the entire spectrum, including the two smeared VH peaks, the low-frequency one being due to transverse modes, and the upper one due instead to the longitudinal mode. The latter contribution, in a more detailed description, would result from hybridization of the longitudinal mode (which lies higher in energy) with a transverse optical (TO) mode, which is not reproduced here in order to keep the model as simple as possible. 

In the bottom panel of Fig.\ref{fig:quattro}, we present the comparison between the model predictions and the experimentally measured VDOS of silica glass under ambient conditions from ~\cite{Monaco2014}. Also in this case, the agreement between theoretical model and experimental data is excellent, and the model is able to reproduce not only the low-frequency part of the spectrum (where the boson peak shows up upon normalizing by the Debye law). Importantly, the theoretical model is able to describe the central part of the spectrum which is entirely dominated by \textit{diffusons}.

In particular, the spectrum of silica appears significantly ''flatter" in comparison with the spectrum of amorphous Si. This is reflected in a larger value of $D_{T}$ for silica glass obtained from the theoretical fit, which signals a stronger extent of disorder in silica compared to amorphous Si. 
In this respect, it is interesting to check the ratio of the values for $D_{T}$ that we have set in $\Gamma_{T}=D_{T} k^{2}$ for the two systems in the fitting, in comparison with independent estimates of the same ratio from simulations and experiments in the literature. 
Taking $D_{T}^{a-Si}$ from simulations of amorphous Si~\cite{Parshin2} and $D_{T}^{silica}$ from acoustic attenuation experiments on silica, focusing on the $\sim k^{2}$ regime~\cite{Baldi2010,Baldi2014}, we obtain the measured value $D_{T}^{silica}/D_{T}^{a-Si}=6.2$, while our theoretical fit gives a value $D_{T}^{silica}/D_{T}^{a-Si}=7.4$, which is a very close value (within error bars of the measured values), and demonstrates the consistency of our theoretical model with the available measurements, and robustness of its predictions.

In summary, we presented a unified theoretical description of the VDOS of amorphous materials in semi-analytical form based on an effective-field theory (EFT) approach to vibrational excitations in solids. Starting from a field-dynamic equation which describes both propagative (\textit{propagons}) transport as well as diffusive scattering (\textit{diffusons}), Eq.\eqref{unob}, a Green's function can be derived with acoustic poles as well as a \textit{diffusive} linewidth due to harmonic ($T$-independent) disorder-induced scattering. This effective description is able to capture all the main features of the VDOS of amorphous materials, including the boson peak (resulting from a Ioffe-Regel crossover between \textit{propagons} and \textit{diffusons}), and the smeared van Hove (VH) peaks, which dominate the upper part of the spectrum. The theoretical model can be used to produce successful fittings of simulated and experimental VDOS spectra of real materials with different chemistry and structure, as demonstrated here on the example of amorphous Si and silica glass. In the future, the model can be employed in a reverse-engineered way to extract dispersion relations of phonons and diffusons, and acoustic attenuation, from measured VDOS  (e.g. Raman~\cite{Kojima1,Kojima2,Ovadyahu}) spectra. This semi-analytic approach opens up new opportunities for the theoretical modelling of thermal properties of solids (thermal conductivity, specific heat) and electron-phonon behaviour. 

\begin{acknowledgments}
\sectionlinetwo{black}{88}
\vspace{0.1cm}

We thank P.B.Allen, G.Baldi, W.Li and G.Ruocco for fruitful discussions and comments on a previous version of this manuscript. M.B. acknowledges the support of the Spanish MINECO’s “Centro de Excelencia Severo Ochoa” Programme under grant SEV-2012-0249.
\end{acknowledgments}
\bibliographystyle{unsrt}
\bibliography{phonon_spectra}

\end{document}